\pdfoutput=1
\documentclass[preprint,showpacs,showkeys,preprintnumbers,aps,amssymb]{revtex4}
\usepackage{graphicx}
\usepackage{epsfig}
\usepackage{amsmath}
\usepackage{bm}
\usepackage[colorlinks=true, pdfstartview=FitV, linkcolor=red, citecolor=blue,
urlcolor=blue]{hyperref}

\def\anp#1#2#3{Annals Phys. {\bf #1}, #2 (#3)}

\def\ibid#1#2#3{{\it ibid.} {\bf #1}, #2 (#3)}

\def\ijma#1#2#3{Intl. Jour. Mod. Phys. A {\bf #1}, #2 (#3)}
\def\jhep#1#2#3{Jour. High Energy Phys. {\bf #1}, #2 (#3)}

\def\npa#1#2#3{Nucl. Phys. A {\bf #1}, #2 (#3)}
\def\npb#1#2#3{Nucl. Phys. B {\bf #1}, #2 (#3)}

\def\prd#1#2#3{Phys. Rev. D {\bf #1}, #2 (#3)}
\def\prl#1#2#3{Phys. Rev. Lett. {\bf #1}, #2 (#3)}

\def\rmp#1#2#3{Rev. Mod. Phys. {\bf #1}, #2 (#3)}

\newcommand{\tr}{{\rm tr}}
\newcommand{\be}{\begin{equation}}
\newcommand{\ee}{\end{equation}}
\newcommand{\bea}{\begin{eqnarray}}
\newcommand{\eea}{\end{eqnarray}}

\newcommand{\p}{\partial}
\newcommand{\s}{\sigma}

\newcommand{\la}{\langle}
\newcommand{\ra}{\rangle}
\newcommand{\rd}{\mbox{d}}

\newcommand{\lqcd}{\Lambda_{{\rm QCD}}}
\newcommand{\Nc}{N_{{\rm c}}}
\newcommand{\Nf}{N_{{\rm f}}}

\renewcommand{\vec}[1]{{\bm #1}}

\newcommand{\arXiv}[1]{\href{http://arXiv.org/abs/#1}{[arXiv:#1]}}

\begin{document}
\title{Covering the Fermi Surface with Patches of Quarkyonic Chiral Spirals}

\author{Toru Kojo$^{a}$, Robert D. Pisarski$^{b}$, and
A. M. Tsvelik$^{c}$}

\affiliation{
$^{a}$RIKEN/BNL Research Center, Brookhaven National Laboratory
Upton, NY-11973, USA\\
$^{b}$Department of Physics, Brookhaven
National Laboratory Upton, NY-11973, USA\\
$^{c}$Department of Physics, Department of Condensed Matter
Physics and Materials Science, Brookhaven National Laboratory,
  Upton, NY 11973-5000, USA\\
}
\date{\today}

\begin{abstract}
We argue that in cold, dense quark matter, in the limit of a large number
of colors the ground state is unstable with respect to creation
of a complicated Quarkyonic Chiral Spiral (QCS) state, in which 
both chiral and translational symmetries are spontaneously broken.
The entire Fermi surface is covered with patches of QCSs,
whose number increases as the quark density does.
The low energy excitations are gapless, given by 
Wess-Zumino-Novikov-Witten model plus transverse kinetic
terms localized about different patches.
\end{abstract}


\maketitle
%
%

\section{Introduction}

In cold, dense quark matter, one expects several phenomenon:
a transition to a deconfined phase,
color superconductivity \cite{color_super}, 
and for a large number of colors, $\Nc$, chiral density waves
\cite{dgr,son,otherA,bringoltz}.

Recent theoretical work has argued that at least for a 
large number of colors, $\Nc$, further phenomena can occur
\cite{McLerran:2007qj,Hidaka:2008yy,Fukushima:2008wg,Andronic:2009gj,Kojo:2009ha,Glozman,guo}.  
There can be a phase in which the baryon density is large, but that is 
nevertheless confining.
While such a ``Quarkyonic'' phase is certainly present at large $\Nc$,
it also exhibits several properties which may have been observed in
experiments at the CERN SPS \cite{Andronic:2009gj}.

Ref. \cite{Kojo:2009ha} suggests that 
in a Quarkyonic phase, the {\it local} breaking of chiral symmetry is driven
by the appearance of chiral density waves, which were termed
Quarkyonic Chiral Spirals (QCS).
This arises from the condensation of particle-hole pairs near the
Fermi surface, and is similar to the
mechanism of chiral condensation in the vacuum.  
The essential difference is that in the presence of a Fermi sea,
the dominant condensation involves a particle-hole pairs
with nearly opposite Fermi momentum.  
The net momentum of the pair is then nonzero and large, so that not
only chiral symmetry, but also translational symmetry, is spontaneously
broken through the formation of a chiral condensate which was termed
a (Quarkyonic) Chiral Spiral.  This condensate rotates between the usual
chiral condensate, $\overline{\psi} \psi$, and a helicity condensate,
$\overline{\psi} \gamma_0 \gamma_z \psi$.  
This is closely analogous to
a chiral spiral in $1+1$ dimensions, which rotates between 
$\overline{\psi} \psi$ and $\overline{\psi} \gamma_5 \psi$
\cite{thies}.

The analysis in Ref. \cite{Kojo:2009ha} was restricted to 
chiral spirals which form in one direction.  
In this paper we argue that the entire Fermi surface is covered with
patches of QCSs, where each patch
corresponds to a different direction for a single QCS.
The long-range forces present in Quarkyonic matter 
imply that the gaps are strongly momentum dependent.
As a consequence, at large angles the interaction between particles and QCSs
is suppressed, and QCSs at large angles to one another interact weakly.

We estimate the width of the patches, and find that
the number of patches increases with the density.
As the density increases, the Fermi surface is characterized by 
increasingly high symmetry, and a series of phase transitions occur.
All of these transitions occur in a confined phase with 
locally broken chiral symmetry, 
and thus are distinct from the transitions
for the restoration of chiral symmetry, or for deconfinement.

Ref. \cite{Kojo:2009ha} showed that QCSs dominate at intermediate
chemical potentials, up to $\mu \sim 80$~GeV.  At higher $\mu$, the effects
of perturbative interactions must also be included, and also generate
chiral spirals.  It was previously suggested in Refs. \cite{son}
and \cite{otherA}
that perturbative chiral spirals also cover the Fermi sphere in patches.  Our
analysis extends and compliments theirs.

\section{The qualitative picture for the confining model}

In the limit of large $\Nc$, the gluon propagator is unaffected by the quarks. 
This means that at low temperature, the gluon propagator 
is the same as in the confined vacuum.
We thus use the effective gluon propagator 
\bea
D_{00}(\omega,{\bf q}) = \frac{\s}{ |{\bf q}^2|^{2}}, ~~ D_{ik}(\omega,{\bf q})
= \frac{\delta_{ik} - q_iq_k/q^2}{\omega^2 + {\bf q}^2}.
\label{confint}
\eea
This propagator is valid in Coulomb gauge, $\partial_i A_i = 0$, 
at small momenta, $|\vec{q}| <\lqcd $.
This propagator is very long range, and 
through the exchange of $A_0$ fields,
strongly enhances the density-density interaction
at small momentum transfer.
This reflects the linear confinement of 
mesonic excitations.  Following \cite{Kojo:2009ha}, we
concentrate on nonperturbative contributions 
from $D_{00}$ and 
ignore perturbative effects from $D_{ik}$.

We summarize the results of \cite{Kojo:2009ha},
and give a qualitative interpretation which we 
use to construct a Fermi surface with multiple patches.
Taking the gluon propagator $D_{00}$ of Eq. (\ref{confint}), the
Schwinger-Dyson equation for the quark self energy can be
constructed, especially for excitations near the Fermi surface. 
It is
\begin{equation}
\not\!\Sigma(p) +\Sigma_m(p)
= - \int \frac{d^4 k}{(2\pi)^4} \;
 (\gamma_0 t_A) \; S(k;\Sigma) \; (\gamma_0 t_B)
\; D^{AB}_{00}(p-k)\;.
\label{SD1}
\end{equation}
Near the Fermi surface, $p_z \sim \mu,$ and $|\vec{p}_T| \sim 0$.
Since the dominant contributions come from interactions with
small momentum transfer, the integral is dominated
by virtual quarks within $\sim \lqcd$ of the edge of the Fermi sea.
In this region the transverse momenta of the quarks can be ignored,
and the integral equation factorizes,
\begin{equation}
\int dk_0 dk_z d^2\vec{k}_T 
\; S(k_0, k_z, \vec{k}_T) \; D_{00}(p-k)\;
\rightarrow 
\int dk_0 dk_z  
\; S(k_0, k_z, \vec{0}_T) 
\int d\vec{k}_T \; D_{00}(p-k) \; .
\label{SD2}
\end{equation}
Integrating over the gluon propagator over $\vec{k}_T$ gives the 
Schwinger-Dyson equation 
for QCD in $1+1$ dimensions in axial gauge, $A_z=0$.

When spatial dimensions are compactified, usually motion in the
small, compact dimensions is energetically disfavored.  In contrast,
here the quarks move easily in the transverse dimension, with essentially
no energetic penalty.  Thus counting the number of states with
different transverse momenta matters, and is taken into account
in the second integral of Eq. (\ref{SD2}).
This smears the four dimensional, confining gluon propagators, 
and reduces them to propagators in two dimensions.

Next consider the particle-hole correlations.
These can be treated in a Bethe-Salpeter equation
for quarks and quark holes, matching north and south poles of
the Fermi sea.  By the same reasoning as applies for the Schwinger-Dyson
equation, the Bethe-Salpeter equation also reduces to that for QCD in
two dimensions.

One might ask why
we only choose states at opposite ends of the Fermi sea, and do not
simultaneously take into 
account contributions from the entire Fermi surface.
For example, in the usual analysis of (s-wave) BCS superconductivity,
the Coulomb force is additive, 
and one electron feels a sum of 
interactions from all electrons over the entire Fermi surface.

This is not true in a Quarkyonic phase.
For a system with quark and quark-holes,
because of confinement what is dominant is the formation of mesons
between quarks and quark holes.  The resulting mesonic states then
interact weakly, via residual meson-meson interactions $\sim 1/\Nc$.
Therefore it is reasonable to first
identify the most tightly bound pairs,
treating interactions between the mesons as a small perturbation.

This is why we first consider a problem with two patches, and then
extend this analysis to multiple patches.
The most relevant pairing for two patches
is the density wave channel in which quarks and quark holes
move in the same direction with total momentum, $\sim 2\mu$,
while keeping their relative momentum to $\sim \lqcd$ \cite{Kojo:2009ha}.
Confinement then drives a gap $\sim \lqcd$, and is not very sensitive
to the quark density.  This will allow us to easily estimate
the transverse size of the patches.

Finally we briefly mention the mapping between quark bilinears
in $3+1$ dimensions to those in $1+1$ dimensions.
The basic point is that
the transverse momenta is negligible 
relative to $\sim \mu$, so
that quantities which couple to transverse
momenta are subleading.
Thus $\gamma_T$ can be dropped,
as can terms which mix spin.  The spin symmetry converts to a
doubled flavor symmetry in $1+1$ dimensions.
For a dictionary of quark bilinear operators, see \cite{Kojo:2009ha}.

\section{The Single Patch Problem: Qualitative Arguments}
In condensed matter physics, usually density waves form because of
a nesting of the Fermi surface, with particle-hole pairs near the
Fermi surface connected by a common wave vector \cite{tsvelik}.
In one spatial dimension, such nesting is automatic, 
because the boundary of the Fermi sea are just two points.
This is not true in higher dimensions.  For quarks in $3+1$ dimensions,
the boundary of the Fermi sea is spherically symmetric, and
nesting cannot occur without deforming the Fermi sphere.

To develop insight into how the chiral condensate forms,
we first assume that the single particle spectrum for quarks
is a well defined notion.  We then apply mean field theory,
and ignore all of the subtleties from confinement.  In the
last part of this section, we argue more carefully, albeit qualitatively,
as to why this might be correct.

We first consider an order parameter with only one Fourier component. 
We use a mean field approximation where  
the chiral spiral acts on quarks like a periodic background,
$\sim \Delta_Q \cos(\vec{Q}\cdot\vec{r})$.  
This problem is mathematically
equivalent to that of noninteracting electrons 
propagating in a crystalline periodic potential. 
The dispersion relations for the particle and hole branches are given by 
  \bea
  E_{\pm} = \frac{\epsilon({\bf p}) + \epsilon({\bf p} + {\bf Q})}{2} \pm
\sqrt{\frac{\epsilon({\bf p}) - \epsilon({\bf p} + {\bf Q})^2}{4} 
+ |\Delta_Q|^2}
\label{E}
  \eea
where $\epsilon({\bf p})$ is the quark energy.
$\Delta_Q$ is the magnitude of the condensate induced by the chiral
spiral waves at a momentum ${\bf Q}$, and 
is treated as a variational parameter.
An optimum $|\vec{Q}|$ should be about $2p_F$, 
so we can safely replace
$\Delta_Q$ with $M \equiv \Delta_{2p_F}$.

Let us establish what the requirements are for ${\bf Q}$ and 
the patch size. 
The $z$-axis is taken to penetrate a center of two patches
connected by $\bf Q$,
and the momenta $\vec{p}$ is measured from $p_F \vec{e_z}$.
Expanding the energies about the Fermi surface, and taking
$Q \sim 2 p_F$, $p_z \sim p_F$,
\bea
&&   \epsilon({\bf p}) = \sqrt{(p_z +p_F)^2 + p_{\perp}^2} 
- p_F \approx p_z + \frac{p^2_{\perp}}{2p_F} + \ldots, \nonumber\\
&& \epsilon({\bf p} + {\bf Q}) 
= \sqrt{(Q - p_z -p_F)^2 + p_{\perp}^2} - p_F
\approx Q- 2p_F - p_z + \frac{p^2_{\perp}}{2p_F} + \ldots \; .
\label{band}
\eea
Substituting this into (\ref{E}) gives
\bea
  E_{\pm} \approx \frac{Q}{2}-p_F + \frac{p^2_{\perp}}{2p_F}
\pm  \sqrt{\left(\frac{Q}{2} - p_F - p_z\right)^2 + M^2}.
\eea

The dispersion relation for a single quark can be used to estimate
the size of the patches.
Energetically, the system benefits from opening as large a gap as possible.
The edge of a patch occurs when a single quark becomes
gapless.  This reflects the balance between 
the condensation energy and the kinetic energy.

Thus $\bf Q$, and the size of the patch, is determined as follows.
A lower bound for $|\vec{Q}|$ arises from the
condition that the positive branch has nonzero energy, $E_+ \geq 0$.
The minimum value of $|Q|$ occurs when
$E_+ = 0$; this happens for
$p_{\perp} =0, p_z = Q/2 -p_F$, which satisfies
  \bea
  E_+^{\min} = \frac{Q}{2} -p_F + M \ge 0.
  \eea
Thus the ordering vector is not exactly $2p_F$,
but a slightly smaller value, $2(p_F-M)$.
The minimum value for $Q$ is $Q_{min} = 2(p_F-M)$,
caused by the deformation of the Fermi surface (see Fig. 1). 

The transverse size of the patches is determined by the condition that
the lower branch has positive energy.
$E_- = 0$ occurs when
  \bea
  p_z = \frac{Q}{2} - p_F=-M\; , \;  ~~ \frac{(p^{\max}_{\perp})^2}{2p_F} = M.
\label{est}
  \eea
For later use, we refer to $p_{\perp}^{\max} = \sqrt{2 p_F M} = M_{\perp}$.

The main origin of the gap is confinement.  In
the Quarkyonic phase at large $\Nc$, the
string tension is independent of density, and thus so is the gap,
 \bea
      M \sim \sqrt{\sigma} \sim \lqcd\; .
 \eea

Why can we use the quark dispersion relation to estimate the transverse
size of the patch?  After all, with a sharp infrared cutoff for 
the confining gluon propagator,
the quark propagator has a pole which diverges as the cutoff vanishes.
This divergence can be avoided with other infrared regulators, but it
is worrisome.

The point is, however, that any quark excitation is inevitably
accompanied by a quark hole, with the two forming a mesonic type
excitation.
For such wavefunctions, infrared ambiguities disappear.
Consider, for example, the theory in vacuum. 
All quarks are confined, but it is sensible to speak of
the constituent mass of a quark, approximately one half the
mass of a meson.  For the mesons near the Fermi surface, then, we can
also speak of a constituent ``mass''.  The dependence of this constituent
mass on the transverse momentum should be given, approximately, by
our naive analysis above.

Of course, the patch size can be rigorously computed by considering
color singlet quantities.  In $1+1$ dimensions, this can be done using
non-Abelian bosonization.  Detailed analysis shows that 
the width of the patch is $\sim p_{\perp}^{\max}$, up to a number of
order one.

%
\section{Multiple Patches}
%

\begin{figure}
\begin{center}
\epsfxsize=0.3\textwidth
\epsfbox{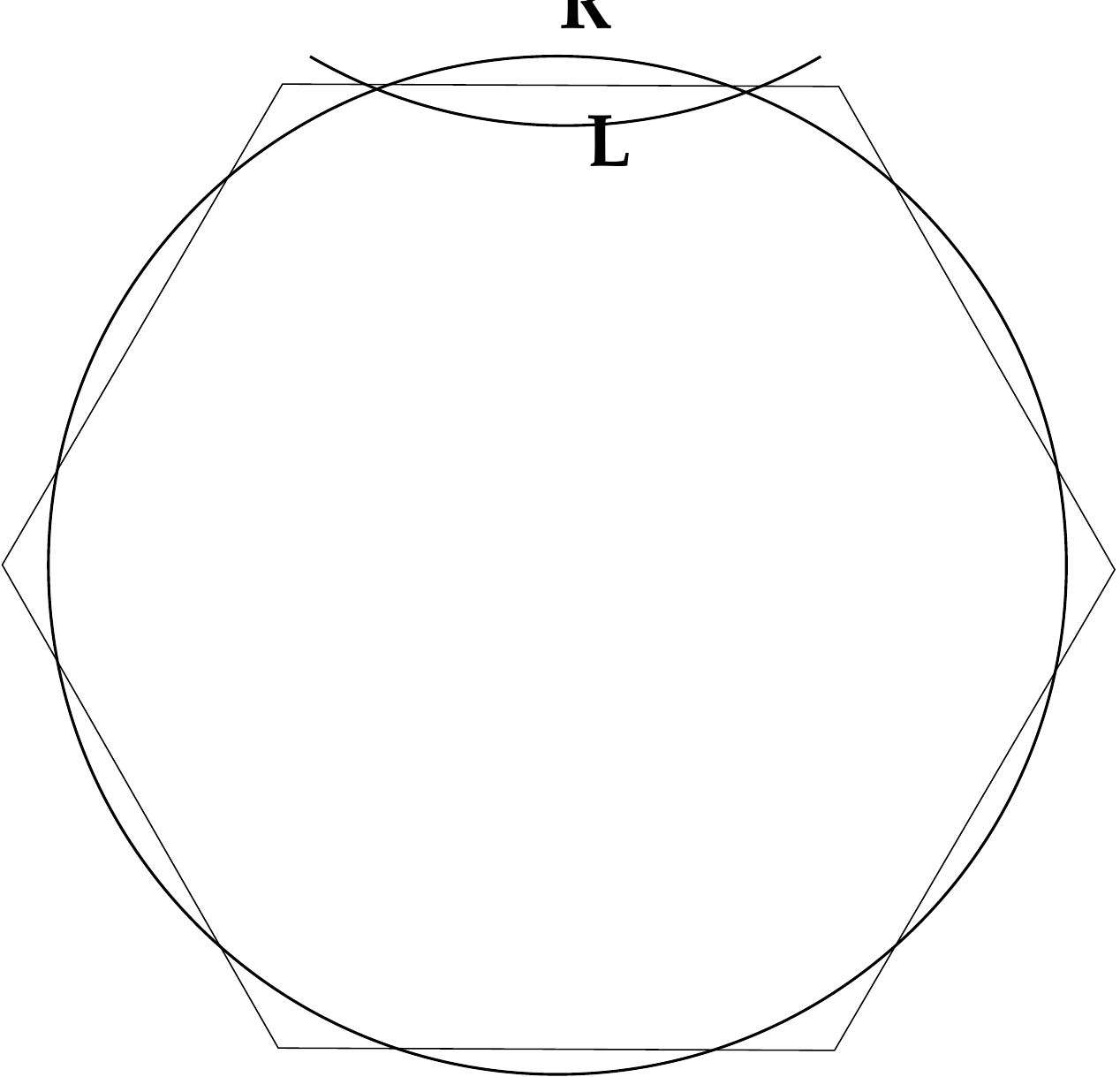}
\end{center}
\caption{The Fermi surface (FS) with three QCSs. 
Bringing the opposite pieces of
FS demonstrates imperfect nesting. 
The spherical  parts of the FS on each edge
are approximated by Eq.(\ref{disp}). }
\end{figure}

In the previous subsection, we considered one patch
problem in $(3+1)$ dimensions. 
The size of the patch was estimated by computing
the transverse momentum at which the gap
closed. 
In reality, the gap remains nonzero, due to interaction with different
patches.  Since the size of the gap is related 
to that of the order parameter, different patches then interact with one
another.

However, a significant simplification takes place when
we take into account that the gap
depends upon the virtuality of the quark.
In vacuum, it is known that
quarks with high virtuality, or in the deep Euclidean region,
do not feel the effects of chiral symmetry breaking.
This suggests that the scattering between a particle and 
a QCS is small if the angle between the two is large.

To see this more explicitly, note that in the presence of a QCS,
the mass self-energy is
\bea
\sum_{i=1}^{N_p} \int \frac{d^4p}{(2\pi)^4}
 \bar{\psi}(p - \vec{Q}_i) M(p;\vec{Q}_i) \psi(p)
+ ({\rm h.c.}).
\label{massvertex}
\eea
where $|\vec{Q}_i| \sim 2p_F$.
Consider the process whereby 
a quark, nearly on-shell 
with momentum $p_0 \simeq p_F$ and $\vec{p}\simeq p_F \vec{e}_z$,
scattering of a QCS with momentum $\vec{Q}_i$, 
and thereby acquires momentum $\vec{p}-\vec{Q}_i$.
QCSs are static, so the energy $p_0$ remains the same.
The mass function $M(p;\vec{Q}_i)$ contributes if the
scattered quark is on-shell within a region $\sim \lqcd^2$,
\bea
p_0^2-(\vec{p} - \vec{Q})^2
\simeq -4 p_F^2 (1-{\rm cos}\theta) \sim - \lqcd^2,
~~~~ \vec{p}\cdot \vec{Q}= 2p_F^2 {\rm cos}\theta.
\eea
Thus the angle to the $z$-axis, $\theta$, is limited to 
$|\theta| 
< \lqcd/p_F$.
Physically, this is the same condition as for the collinear
scattering of two quarks.  This condition is special to the
long range interactions in Quarkyonic matter.

Hence the mass term in (\ref{massvertex})
is dominated by patches whose angle is within $0$ or $\pi$
of the incident quark,
\bea
\sum_{i=1}^{N_p} \int \frac{d^4p}{(2\pi)^4}
 \bar{\psi}(p - \vec{Q}_i) M(p;\vec{Q}_i) \psi(p)
\times 
\theta 
\bigg( \lqcd^2 - |2\vec{p}\cdot \vec{Q}_i - \vec{Q}^2_i| \bigg)
+ ({\rm h.c.}).
\eea
In the following, we will take into account
interactions between nearest neighbor patches in momentum space.

From Eq.(\ref{est}), the number of patches is 
\be
N_p \approx \frac{4p_F^2}{M_{\perp}^2} 
=\frac{p_F}{2M} \sim \frac{n^{1/3}}{M} \; ,
\ee
where $n$ is the quark density per single flavor.
Thus the number of patches increases slowly with the density.
Since the number of patches is an integer, it must change
discontinuously.
In Fig. 2 we depict a succession of four different 
coverings of a sphere, corresponding to different arrangements of QCSs.
At the smallest density there are three mutually 
perpendicular QCSs, which form a cube.
As the density increases, extra patches emerge at the corners of
the cube, giving rise to four more QCSs, the blue shape in Fig.2. 
The wave vectors of the new QCSs are aligned along the diagonals of
the cube, so that eight new patches emerge at the corners
of the cube.  Increasing the density further leads to extra
patches at the corners of the blue shape, leading to the red shape, thence to
the green shape, and so on.

\begin{figure}
\begin{center}
\epsfxsize=0.4\textwidth
\epsfbox{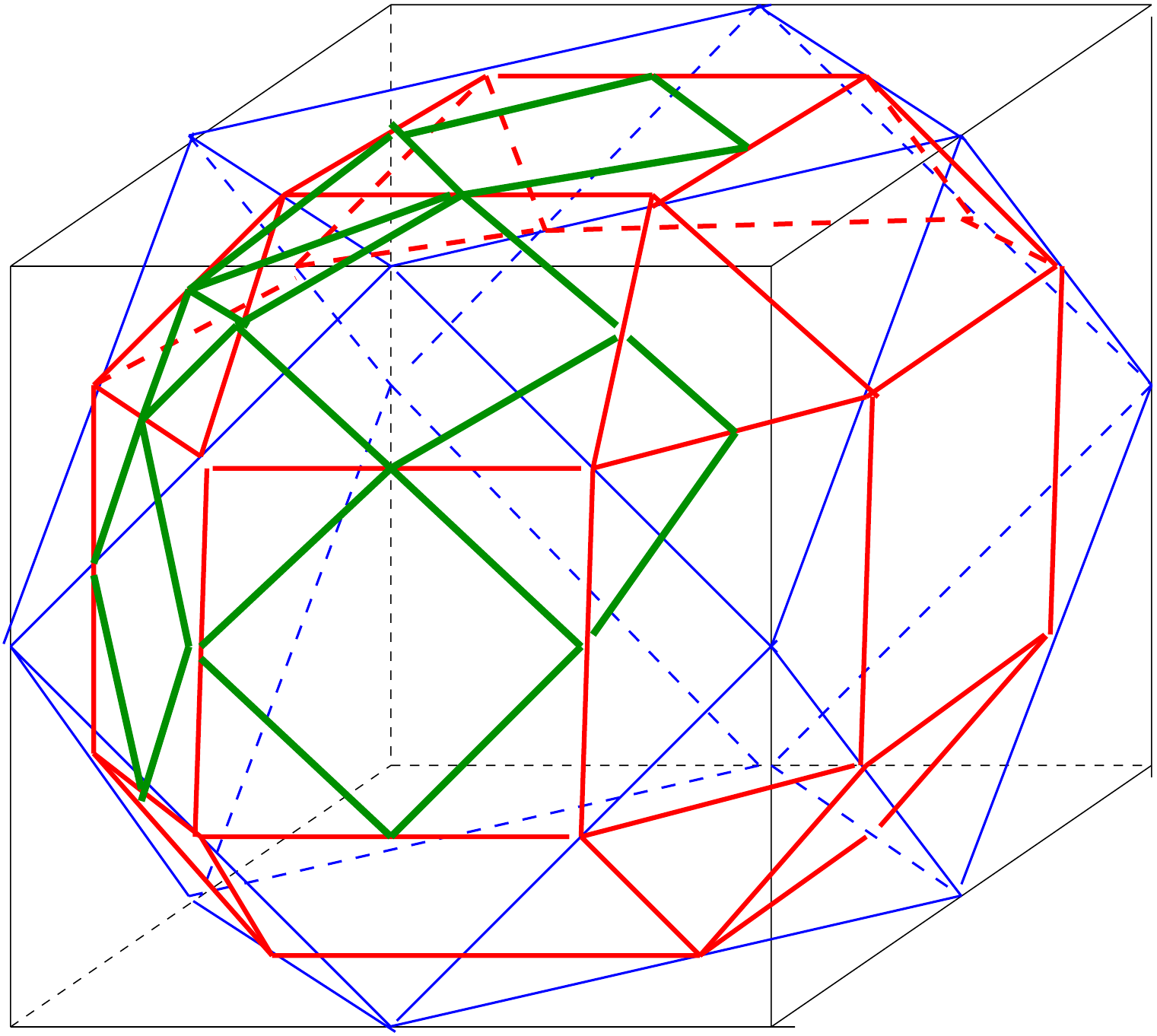}
\end{center}
\caption{A succession of QCS configurations corresponding to the different 
quark density. Black lines show a simple cube.}
\end{figure}

Instead of the above construction, what if one takes 
smaller patches, say of size $\lqcd$, with a corresponding increase
in the number of the patches?
This is not possible, though.  Because the gap arises from
a confining force, the size of $Q$ is fixed
to be $|Q|/2 \sim p_F-M$.  By Luttinger's theorem, the
total volume of the Fermi sea is conserved, so any volume squeezed in the 
$p_z$-direction 
must be compensated by an extension in the transverse size,
until it reaches $\sim M_{\perp}$.
We expect that the presence of 
other patches slightly reduce the transverse size,
$\alpha M_{\perp}$ ($\lqcd/M_\perp \ll \alpha<1$), while
$|\vec{Q}|$ slightly increases from the value for one patch.
Below we will determine $\alpha$ and $|\vec{Q}|$ by regarding
them as variational parameters.

To match different patches onto one another, 
consider the part of the Fermi surface where several patches meet.
From Fig. 2 it follows that the most typical configuration is where
four patches meet at one corner.
We want to establish the conditions for the emergence of next patch.  
Assuming that there are many patches, the curvature at the corners
can be neglected.
We also neglect the influence of all other patches except for those
four which are adjacent.  
Then the problem reduces that of a particle subject to 
the  potential generated by four QCSs.  If $\widetilde{\vec{Q}}$
is a vector in the direction of the corner, then the four
wave vectors of the patches are
${\bf Q}_i = \widetilde{\vec{Q}} \pm \alpha M_{\perp} {\bf e}_i$, where
\bea
\sum_i {\bf e}_i =0, ~~ \vec{e}_i^2=1, 
~~ \widetilde{{\vec Q}}\cdot \vec{e}_i =0, 
~~M_{\perp} = \sqrt{|\widetilde{{\vec Q}}|M}
  \ll  | \widetilde{{\vec Q}}| \; ,
\label{Cond}
\eea
with $M_{\perp} = \sqrt{2 p_F M}$ is the transverse size of the patches,
from above.  We expect that having several patches
does not change this, so we take $\alpha \sim 1$. 
In high density, 
we can set $\widetilde{\vec{Q}}\simeq \vec{Q}$ with $|\vec{Q}|=2p_F$,
and ${\vec Q}\cdot \vec{e}_i \simeq 0$.

Near the corner, the total gap is
\bea
\Delta_Q(\vec{r}) \simeq
\frac{M}{4}
 \sum_{i} e^{i(\vec{Q} \pm \alpha M_{\perp} {\bf e}_i)\cdot \vec{r} }
= \frac{M e^{iQz}}{2}
  \Big[\cos (\alpha M_{\perp}x) + \cos(\alpha M_{\perp}y) \Big]
\; .
\eea
By the previous arguments about virtuality,
only particles within $\sim \lqcd$ of a corner contribute.
Away from the corner, where $|\vec{p}_\perp| \gg \lqcd$,
we reduce to the previous problem of one patch.

Since the transverse modulation occurs slowly,
we can calculate the spectrum in the adiabatic approximation,
separating the motion along ${\bf Q}$ from the transverse motion. 
Initially we consider the transverse coordinates as constant parameters.
Then the energy eigenstates yield the two bands 
of the one patch problem, Eq. (\ref{band}).
Next we consider the energy for the
effective Hamiltonian, $H_{eff}^{\pm}$, describing 
the transverse motion of the upper ($+$) and lower ($-$) bands, 
\bea
&& H_{ {\rm eff} }^{\pm} 
\simeq \left(\frac{Q}{2}-p_F \right)- \frac{1}{Q}(\nabla_{\perp})^2 
 \pm  \sqrt{\left(k_z - \frac{Q}{2}\right)^2 
 + |\Delta({\bf r}_{\perp})|^2} \; , \nonumber\\
&& |\Delta({\bf r}_{\perp})|^2 
= 
 M^2 \Big[\cos (\alpha M_{\perp}x) + \cos(\alpha M_{\perp}y )\Big]^2 \; .
\label{bands}
\eea
This patch configuration becomes unstable 
when the gap between the upper and the lower bands close. 
This occurs first when $k_z = Q/2$. Eq. (\ref{bands}) 
is the Schr\"odinger equation in a periodic potential. 
Its solutions and eigenvalues $E_l({\bf p}_{\perp}),\ l=1,2,...$,
depend on the continuous variable ${\bf p}_{\perp}$ (quasi-momentum) 
with values confined to the Brillouin zone 
$-\alpha M_{\perp}/2 < (p^x_{\perp},p^y_{\perp}) < \alpha M_{\perp}/2$.
The solutions are arranged so that $E_1 < E_2 < E_3 <...$. 
The given QCS configuration becomes unstable 
if the lowest band, $l=1$, crosses from negative to positive energy
at some value of quasi-momentum. 

Taking into account (\ref{Cond}) and making the substitution 
${\bf r}_{\perp} \rightarrow (\alpha M_{\perp})^{-1}{\bf r}_{\perp}$, 
we get the Schr\"odinger equation for the transverse motion:
\bea
&& 
\left(- \alpha^2(\p_x^2 + \p_y^2) \pm \frac{|\cos x + \cos y|}{2}
+ \frac{1}{M} \left(\frac{Q}{2}-p_F \right) \right)\Psi 
= \left(\frac{E_{\pm}}{M}\right)\Psi, \nonumber\\
&& 
\Psi({\bf r}_\perp + 2\pi{\bf a}) 
= {\rm exp} \left( 
\frac{2\pi i ({\bf p}_{\perp}\cdot {\bf a}) }
{\alpha M_\perp} \right) \Psi({\bf r})
\eea
where ${\bf a} = (n,m)$ is a lattice vector. 
Remember that $Q/2-p_F<0$, so
positive and negative branch conditions are not automatically satisfied.
We have to choose appropriate regions for $\alpha$ and $|\vec{Q}|$. 

Let us find such a region.
The transverse equation can be solved analytically 
at small and large $\alpha$. 
At $\alpha < 1$, we can approximate the potential as 
a sum of harmonic wells.  For the negative branch, 
\bea
\frac{E_-}{M} &=& \frac{1}{M}\left(\frac{Q}{2}-p_F\right)
-2 + 2\alpha - \frac{\alpha^2}{4} + O(\alpha^3) \nonumber\\
&& 
- \frac{1}{2} 
  \left( \cos\left(\frac{2\pi p_x}{\alpha M_\perp}\right) 
+ \cos\left(\frac{2\pi p_y}{\alpha M_\perp}\right) \right)
~\exp\left(- \frac{\pi^2}{2\alpha}\right)
\label{disp0}
\eea
Since $(Q/2-p_F) \sim -M$, the last term is
small.  At small $\alpha$, $E_0 \approx -3 + O(\alpha)$, and satisfies
the negative branch condition. 

At large $\alpha \gg 1$, the spectrum $E_-$ is dominated 
by the kinetic energy, $E_- \sim \alpha^2$, and is positive. 
Thus the spectrum crosses at some positive value of $\alpha \sim 1$,
in agreement with the rough estimates of Eq. (\ref{est}).
When $\alpha$ reaches the critical value,
the gap between the bands shrinks to zero at some point 
in the Brillouin zone, creating conditions 
for the emergence of new QCS with a wave vector at this point.  

\section{Effective action for the collective excitations.}

We now derive the  effective action for collective excitations. This is
possible to do even without detailed knowledge of the quark correlation
functions. We begin with the collective modes located
on a particular patch, treating the curvature of the Fermi surface as a
perturbation.  

Consider two opposite  patches of the Fermi surface
connected by vector ${\bf Q}$, taking the $z$-axis along 
${\bf Q}$.  
Diagonalizing the Dirac Hamiltonian for quarks 
close to the Fermi surface,
 \bea
 H_0 & = & 
\int dz d^2\vec{x}_\perp 
\Big\{ i (-R^\dagger\p_zR + L^\dagger\p_z L) 
+ \frac{1}{2p_F}(\nabla_{\perp}R^\dagger\nabla_{\perp}R +
\nabla_{\perp} L^\dagger\nabla_{\perp}L) + \cdots \Big\} \; ;
 \eea
where $R,L$ fields represent right-moving and left-moving particles
along $z$-direction.
We suppress all indices, including those for spin.
We consider $\Nf$ flavors in $3+1$ dimensions, so that 
fermion fields in one-dimensional chain effectively have $\Nf'=2\Nf$ 
flavor indices.

We treat the system as many one-dimensional chains
which couple through transverse kinetic terms.
We discretize the transverse dimensions, so that the
transverse degrees of freedom emerge as a result of interchain tunneling
\cite{essler},
 \bea
 && 
H_0 = \sum_{\bf r} \int\rd z \Big\{i 
\big(-R^\dagger({\bf r})\p_z R({\bf r}) 
+ L^\dagger({\bf r})\p_z L({\bf r}) \big) \nonumber\\
 && 
\hspace{2cm}
- t\sum_{{\bf a}} \Big[R^\dagger({\bf r}) R({\bf r} +{\bf a}) 
+ L^\dagger({\bf r})L({\bf r} +{\bf a})\Big]\Big\} \; .
 \eea
Here $\vec{a}$ are vectors connecting nearest neighbors, and 
$t$ and $a$ are chosen
so that the spectrum of the discrete model approximates the
continuous theory. The requirement of nearest neighbor tunneling is not 
necessary, and can be generalized, if need be.
Choosing a square lattice, the resulting dispersion relation is
 \bea
\epsilon_{\perp}(\vec{p}_\perp) 
 = -2t\sum_{\vec{a}} \cos[\vec{a}\cdot \vec{p}_\perp ] 
=-2t[\cos(p_x a) + \cos(p_y a) - 2] 
\label{disp}
 \eea
In the continuum limit this dispersion relation reproduces
$\vec{p}_\perp^2/2p_F$ on the interval set by (\ref{est}):
 \bea
 t a^2 = 1/2p_F, ~~ 8t = M. \label{ta}
 \eea
Therefore $a^{-1} = (p_F M)^{1/2}/2 \sim M_\perp$.
As we have shown in the previous subsection, the transverse dispersion
can be substantially renormalized by the interaction between the patches,
Eq. (\ref{disp0}). 

The Hamiltonian for each patch is ($\Phi=R+L$)
\bea
  && H = \sum_{\vec{r}} H_{1D} (\vec{r}) + T_{tunn}\\
  && H_{1D} = \int dz
\Big\{ i \big( -R^\dagger(\vec{r})\p_z R(\vec{r}) 
+ L^\dagger(\vec{r})\p_z L(\vec{r}) \big)  \nonumber \\ & &  
\hspace{3cm} 
+ \int dw \; \frac{g_{\rm 2D}^2}{2} 
\big( \Phi^\dagger t_A \Phi (\vec{r}, z) \big)
\; |z-w| \; 
\big( \Phi^\dagger t_A \Phi (\vec{r},w) \big) \Big\} \label{1D}\\
  && T_{tunn} = - t\sum_{\vec{r}, \vec{a}}
\Big[R^\dagger(\vec{r}) R(\vec{r} + \vec{a}) +
L^\dagger(\vec{r})L(\vec{r} + \vec{a})\Big]\label{tunn}
\eea
It is important that the interaction is essentially one-dimensional and is
concentrated inside of each chain. 
One-dimensional models (\ref{1D}) can be
treated by non-Abelian bosonization, as was done in \cite{Kojo:2009ha}.  
The long range
current-current interaction generates a gap of magnitude $M$ in the color
channel.  The remaining gapless modes are 
described by the $U_{\Nf' \Nc}(1)$ and
$SU_{\Nc}(\Nf')$ Wess-Zumino-Novikov-Witten (WZNW) models. 

Some care must be taken in the normalization of quark fields,
originally in $3+1$ dimensions, when we apply bosonization rules
to the corresponding fields in $1+1$ dimensions.
A one dimensional chain has resolution $\sim a^{-1} \sim M_\perp$ 
in the transverse direction, so it is necessary to multiply 
dimensionless fields by $a^{-2}$ when we include
the transverse dimensions, $z \rightarrow (z,\vec{r})$.  For instance,
\bea
\bar{\Phi} \Phi (z,\vec{r}) 
\rightarrow a^{-2} \times C_{1D} ( U \; \tr[g] \; \tr[h] + (h.c.) )
\eea
where the matrix valued fields $U$, $g$, and $h$ on the right hand side
are boson fields for $U(1)$, flavor, color sector.
All of these fields are dimensionless.
From the analysis of QCD in $1+1$ dimensions,
$C_{1D}$ has dimensions of mass, $\sim \lqcd$.

Next we treat Eq. (\ref{tunn}) as a perturbation on the WZNW models.
Turning on the transverse interactions, two things happen.
While the single particle gap closes at the border of each patch,
as discussed in the previous section, the gap is restored when
we take into account the fact that several patches meet at that point.
In the mean field approximation, the size of the patch is given by
Eq. (\ref{est}). A more rigorous approach would be
to use the exact 1D quark's Green's function for $1+1$ dimensional QCD,
and to use the 
Random Phase approximation (RPA) expression for the single quark Green's
function, in the spirit of Ref. \cite{essler}:
 \bea
 G_{RR} = [G_{RR,1D}^{-1}(\omega,p_z) - \epsilon({\bf p}_{\perp})]^{-1},
\label{green}
 \eea
where $G_{RR,1D}$ is the Green's function for right-moving quarks in the
one dimensional Hamiltonian of Eq.(\ref{1D}). 
A similar expression, with $R$ replaced by $L$, holds for
the left-moving quarks. Then the size of the patch is determined by the
transverse momentum at which the Green's function (\ref{green}) acquires a
singularity at $\omega =0$:
 \bea
 \frac{p_{\perp}^2}{2p_F} = \mbox{min}\; G_{RR,1D}^{-1}(0,p_z)
 \eea
At present we do not know the form of $G_{1D}$; if we assume that
 \bea
 G_{1D}(0,p_z) = \frac{1}{p_z} \; f\left(\frac{p_z^2}{M^2}\right)
 \eea
then we find that
 \bea
\frac{(p^{\max}_{\perp})^2}{2p_F} = \beta M \; , ~~ \beta \sim 1 \; .
 \eea
This agrees  with the mean field estimate of Eq. (\ref{est}).

The second effect of the interchain tunneling is that it generates a coupling
between collective modes inside of a given patch. This occurs through virtual
scattering processes, into states above the single particle gap.
In the interaction picture for $H_{1D}$, 
time ordered product in second order in $t$
generates the interchain coupling,
\bea
&& t^2
 \sum_{\vec{r},\vec{r'}, \vec{a}, \vec{a'} } \int d^2z d^2w 
\big[R^\dagger(z, \vec{r}) R(z, \vec{r} + \vec{a}) \big] 
 \big[ L^\dagger(w,\vec{r}'+ \vec{a}') L(w, \vec{r}') \big]
\nonumber \\
&& =
- t^2 a^2
 \sum_{\vec{r}, \vec{a} } \int d^2z d^2w
\big[R_{fc}^\dagger(z, \vec{r}) L_{f'c'}(w, \vec{r}) \big]
 \big[ L_{f'c'}^\dagger(w,\vec{r} + \vec{a}) 
R_{fc} (z, \vec{r} + \vec{a}) \big]
+ \cdots  \; ,
\label{trans}
\eea
where $f,c$ express flavor and color indices, respectively.
We neglect terms where $\vec{r} \neq \vec{r}'$
or $\vec{a} \neq \vec{a}'$, which do not contribute
at this order.

We integrate over the color sector, leaving 
color singlet operators.
We use coherent field expressions, 
$L = (a^{-1} C_{1D}^{1/2}) \; \xi_L \xi_{Lf} \xi_{Lc}$ 
and $R = (a^{-1} C_{1D}^{1/2}) \; \xi_R \xi_{Rf} \xi_{Rc}$,
which satisfy 
$\xi_R^\dag \xi_L = U$, $\xi_{Rf}^\dag \xi_{Lf'} = g_{ff'}$, 
and $\xi_{Rc}^\dag \xi_{Lc'} = h_{cc'}$.
Due to $U(1)$-flavor-color separation in $H_{1D}$,
we can evaluate each sector separately.
Note that only color sector has the scale $\sim \lqcd \sim M$.
Further, correlation functions in the color sector damp rapidly,
while those for the $U(1)$ and flavor sectors have power law correlations.
This scale hierarchy allows us to use the following approximation:
\bea
\la R_{fc}^\dag(z) L_{f'c'}(w) \ra^{color}_{H_{1D}}
&=& \frac{C_{1D}}{a^2} \;
\big( \xi^\dag_R(z) \xi_L(w) \big)\;
\big( \xi^\dag_{Rf}(z) \xi_{Lf'}(w) \big)\;
 \big\la \xi^\dag_{Rc}(z) \xi_{Lc'}(w) \big\ra_{H_{1D}} 
\nonumber \\
&\simeq&  \frac{C_{1D}}{a^2} \;
\big( \xi^\dag_R(z) \xi_L(z) \big) \;
\big( \xi^\dag_{Rf}(z) \xi_{Lf'}(z) \big) 
D_{cc'} (z-w)
\nonumber \\
&\simeq& 
\frac{C_{1D}}{a^2} \;
\Delta_{ff'}(z) \; 
\big(\delta_{cc'} g (z-w) + (t_A)_{cc'} f_A (z-w) \big) ,
\eea
where $\Delta_{ff'} = 
: e^{i\Phi} :\; :g_{ff'}: $.
Therefore Eq.(\ref{trans}) becomes
\bea
\frac{\gamma \Nc}{M^2} 
\times \bigg(-\frac{t^2 C_{1D}^2}{a^2} \bigg)
 \sum_{\vec{r}, \vec{a} } \int d^2z \;
\Delta_{ff'}({\bf r},z)\Delta_{f'f}^\dagger({\bf r} + {\bf a},z) 
+ O(\Nc^0) \; .
\eea
The first factor arises from integration of 
correlation functions in color sector,
and $\gamma \sim 1$. 
We have used the fact 
that correlation functions in the color sector contribute
to integrals only within a distance $\sim M^{-1}$.
Corrections $O(\Nc^0)$ arise from color non-singlet operators.

We now have the effective Lagrangian for interchain coupling:
\bea
L_\perp  
= \frac{\gamma \Nc C^2_{1D}}{(8a)^2} \sum_{{\bf r,a}}\int d^2z \;
 \tr [\Delta({\bf r},z)\Delta^\dagger({\bf r} + {\bf a},z) ]
= \frac{\gamma \Nc C^2_{1D}}{64} 
\int d^4 x \;
\tr[\partial_\perp \Delta(x) \partial_\perp \Delta^\dagger(x)]
\; ,
\label{V}
 \eea
where we used $8t = M$.
For the unperturbed Hamiltonian,
the correlation function of $\Delta$ scales as 
 \bea
 \la \Delta(\tau, z)\Delta(0,0)\ra_{H_{1D}} 
= \Big[\frac{M^{-2}}{\tau^2 + z^2}\Big]^{-d_{\Delta}} \; .
\eea
Here $M$ serves as the ultraviolet cut-off. 
The scaling dimensions of the
primary fields in the WZNW model are \cite{knizhnik}:  
\be
d_{\Delta} = \Big[\frac{1}{\Nc \Nf'} 
+ \frac{\Nf'- 1/\Nf'}{\Nc + \Nf'}\Big] \; . \label{d} 
\ee
Therefore the scaling dimension of the 
operator in Eq. (\ref{V}) is less than two: that is,
transverse hopping has generated relevant operators.

Near zero temperature, the system exhibits long range correlations.
The massless modes inside of each patch couple strongly to one another,
with an energy which depends on their transverse momenta.
Their three dimensional dynamics are described by the effective Lagrangian
\bea
 && {\cal L}_{k=\Nc \Nf'}^{U(1)} 
= \frac{\Nf' \Nc p_F M}{8} 
\Big[\; (\partial_L \Phi)^2
+ \frac{\eta M}{p_F}  (\partial_\perp \Phi)^2  \; \Big] \; ,
\nonumber \\
&&
{\cal L}_{k=\Nc}^{SU(\Nf')}
= \frac{\Nc p_F M}{4} 
\Big[\; {\cal L}_{WZW}[g] 
  + \frac{\eta' M}{p_F} \tr[\partial_\perp g \partial_\perp g^\dag ] 
\; \Big] \; ,
\label{eff}
\eea
where 
\bea
{\cal L}_{WZW}[g]
= \frac{1}{16\pi} \tr[ \partial_L g \partial_L g^\dag ]
+i\frac{\epsilon_{\mu \nu \lambda}} {24\pi}
 \int d\xi \; 
\tr[(g^\dag \partial^\mu g )(g^\dag \partial^\nu g )
(g^\dag \partial^\lambda g)] \; .
\eea
Here $\partial_L$ express derivatives for 
time and longitudinal directions.
We have omitted corrections $O(M^2/p_F^2, 1/\Nc)$, 
with the constants $\eta,\ \eta' \sim 1$.

In one spatial dimension, due to infrared fluctuations 
there is only quasi long range order.  After including transverse
hopping, our gapless modes become three dimensional, Eq. (\ref{eff}),
which drastically decreases the role of infrared fluctuations.
Consequently, in three dimensions there is true long range order,
and a well defined ground state.
(In contrast to Ref. \cite{baym}, this is only seen when
transverse fluctuations are included.)

Another consequence of the increase in the
spatial dimensionality is that nonlinear
terms in the WZNW action can be treated perturbatively.  
Goldstone modes of a QCS have a spectrum
\bea
 \omega_{\bf n}^2({\bf p}) 
= ({\bf n \cdot p})^2 + \kappa \left(\frac{M}{p_F}\right)[{\bf n}
 \times {\bf p}]^2 \; , \; ~~ \kappa \sim \eta,\ \eta' \; .
\eea
The unit vector ${\bf n}$ is the wave vector of the QCS, with
momenta measured from north or south pole of the 
Fermi surface, $p_F \vec{n}$.
This expression is valid for $|\vec{p}_\perp| \ll M_\perp$.
It is not valid close to the boundary of a patch,
where one cannot expand in $t$, and 
effects from other patches must be included.
This shows that at high density the spectrum is less sensitive to 
to the transverse momenta, in accord with Ref. \cite{Kojo:2009ha}.

Finally we mention interactions among quantum excitations.
Since Eq.(\ref{eff}) has a decay constant of $O(\Nc)$,
higher order couplings of quantum excitations always
accompany suppression factor of $O(1/\Nc)$,
in a similar way as chiral Lagrangian.
Therefore, in large $\Nc$,
dynamics close to the Fermi surface is described by
free Goldstone bosons.  The interaction between coherent
excitations are interesting and will be discussed elsewhere.

\begin{figure}
\begin{center}
\epsfxsize=0.4\textwidth
\epsfbox{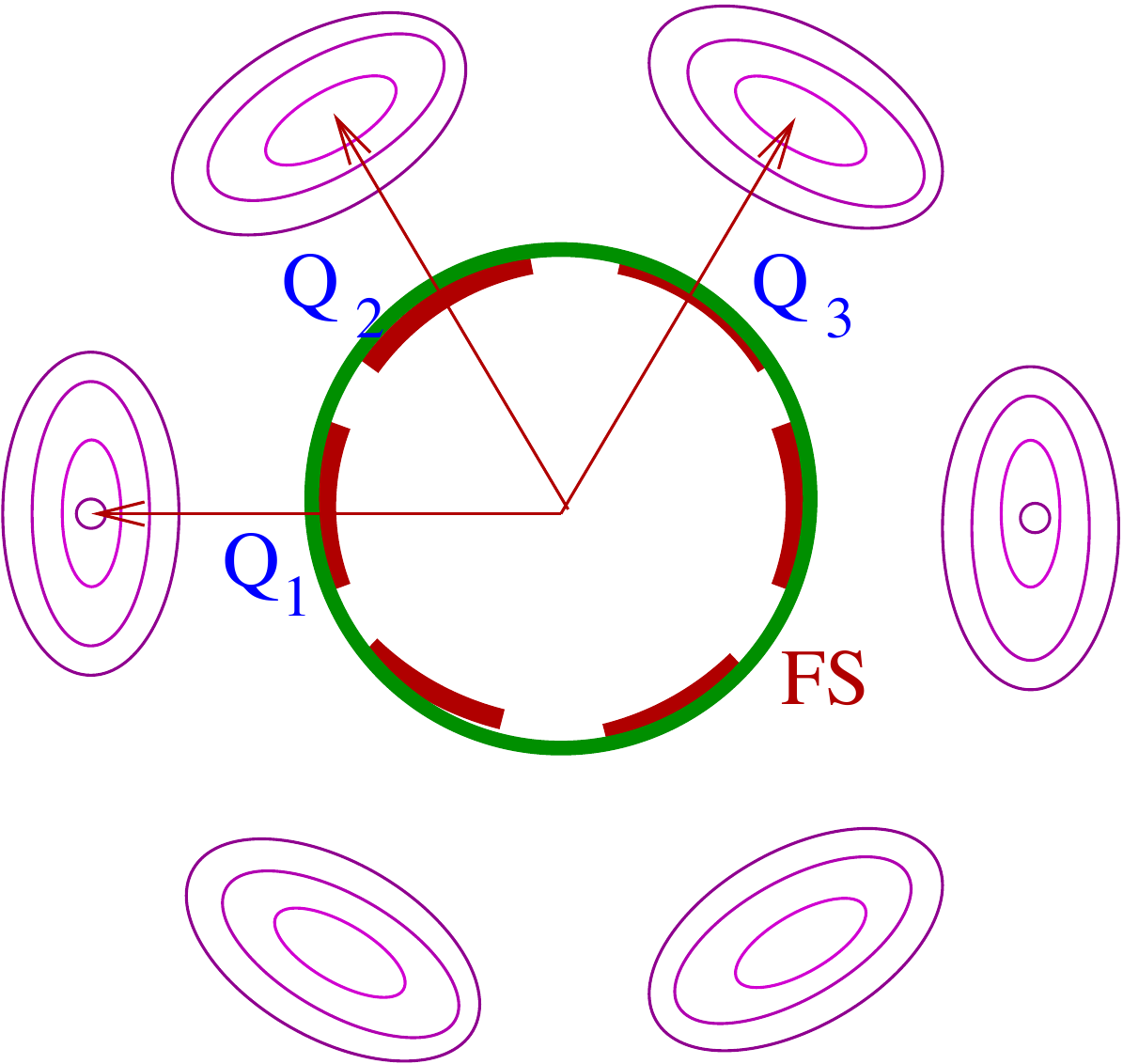}
\end{center}
\caption{The schematic picture of collective excitations for the Fermi surface
with three QCSs. The magenta ellipses of various intensity  correspond to
surfaces of constant $\omega({\bf p})$. The dark areas represent regions of
momentum space where the Goldstone modes are subject to decay into quarks. The
green circle is the original FS. The red arcs represent 
the portions of FS where spectral gaps emerge.  }
\end{figure}
%
\section{Conclusions}
In a Quarkyonic regime, where quarks are confined,
instead of treating the entire Fermi surface, it is more natural
to first construct mesonic degrees of freedom within a particular
momentum region.  Since gaps arise from confinement,
the gap energy is almost independent of density.

We analyzed multiple patches of QCSs based on results for a single patch.
In going from one to multiple patches, it is essential to remember
that interactions between particles (or holes) and QCSs are suppressed
at large angles.
Then each patch can be treated incoherently,
except at boundaries where several patches meet.
Knowing the magnitude of the gap allowed us to estimate the size of each
patch.

We derived the effective Lagrangian for 
Goldstone modes by regarding
the system as one dimensional chains with
weak transverse tunnelings.
After inclusion of transverse hopping effects,
the spectrum of Goldstone bosons propagate in three dimensions.
As a consequence, quasi-long range order in $1+1$ dimensions becomes,
in $3+1$ dimensions, true long range order.
Corrections can be systematically estimated through
an expansion in $\lqcd/p_F$.

Our arguments imply that as the quark density increases, 
the Fermi surface acquires more patches, which have an
increasingly large symmetry.
Accordingly, there is a series of phase transitions as the symmetry
of the Fermi surface increases.
At the same time, the Goldstone bosons become more like those
in $1+1$ dimensions.
Then long range order in $3+1$ dimensions becomes closer to
quasi-long range order in $1+1$ dimensions.
This suggests that both chiral and rotational symmetries are
effectively restored, but in a manner rather different from common
assumption.  

\section{Acknowledgments}
We are grateful to Y. Hidaka, L. McLerran, A. Rebhan,
and A. Schmitt  for discussions and interest in our work. 
The research of R. D. Pisarski and A. Tsvelik are supported under 
the US Department of Energy Contract No. DE-AC02-98CH10886.
T. Kojo is supported by the Special Posdoctoral Research Program of RIKEN.
R. D. Pisarski also thanks the
Alexander von Humboldt Foundation for their support.


\end{document}